\documentclass[aps,prl,twocolumn,superscriptaddress,showpacs]{revtex4} % UTOBS
\newcommand{\figsize}{0.4\textwidth}      %              UTOBS
\newcommand{\figsizes}{0.35\textwidth}      %              UTOBS
\usepackage{graphicx}% Include figure files
\usepackage{dcolumn}% Align table columns on decimal point
\usepackage{bm}% bold math
\usepackage{setspace}
\def\msq{M_{\widetilde q~}}             %squark mass
\def\mgl{M_{\widetilde g~}}             %gluino mass

\def\met{\mbox{${\hbox{$E$\kern-0.6em\lower-.1ex\hbox{/}}}_T$}} %missing ET
\def\mex{\mbox{${\hbox{$E$\kern-0.6em\lower-.1ex\hbox{/}}}_x$}} %missing Ex
\def\mey{\mbox{${\hbox{$E$\kern-0.6em\lower-.1ex\hbox{/}}}_y$}} %missing Ey
\def\mexy{\mbox{${\hbox{$E$\kern-0.6em\lower-.1ex\hbox{/}}}_{x,y}$}} %missing Exy
 
\def\D0{D\O}                            %D0
\newcommand{\Dzero}{\rm{D0}}

\newcommand{\etjet}{E_{  T}^{\rm jet}}
\newcommand{\etjetem}{E_{  T,{\rm em}}^{\rm jet}}
\newcommand{\etjeta}{E_{ T}^{\rm jet(1)}}
\newcommand{\etjetb}{E_{ T}^{\rm jet(2)}}
\newcommand{\etjetc}{E_{ T}^{\rm jet(3)}}
\newcommand{\etjetd}{E_{ T}^{\rm jet(4)}}
\newcommand{\etjeti}{E_{ T}^{\rm jet(i)}}
\newcommand{\etajet}{\eta^{\rm jet}}
\newcommand{\ptch}{p_{T,\rm trk}^{\rm jet}}

\def\tt{\mbox{$t \overline{t}$}}
\begin{document}
%
% ======> Title of the paper goes here <====================
%
\title{
Inclusive Search for Squark and Gluino Production  in 
{\boldmath $p\overline{p}$} Collisions at 
{\boldmath $\sqrt{s}$} = 1.96 TeV \\
%(version v1 \ \today)
}

\affiliation{Institute of Physics, Academia Sinica, Taipei, Taiwan 11529, Republic of China} 
\affiliation{Argonne National Laboratory, Argonne, Illinois 60439} 
\affiliation{University of Athens, 157 71 Athens, Greece} 
\affiliation{Institut de Fisica d'Altes Energies, Universitat Autonoma de Barcelona, E-08193, Bellaterra (Barcelona), Spain} 
\affiliation{Baylor University, Waco, Texas  76798} 
\affiliation{Istituto Nazionale di Fisica Nucleare Bologna, $^w$University of Bologna, I-40127 Bologna, Italy} 
\affiliation{Brandeis University, Waltham, Massachusetts 02254} 
\affiliation{University of California, Davis, Davis, California  95616} 
\affiliation{University of California, Los Angeles, Los Angeles, California  90024} 
\affiliation{University of California, San Diego, La Jolla, California  92093} 
\affiliation{University of California, Santa Barbara, Santa Barbara, California 93106} 
\affiliation{Instituto de Fisica de Cantabria, CSIC-University of Cantabria, 39005 Santander, Spain} 
\affiliation{Carnegie Mellon University, Pittsburgh, PA  15213} 
\affiliation{Enrico Fermi Institute, University of Chicago, Chicago, Illinois 60637} 
\affiliation{Comenius University, 842 48 Bratislava, Slovakia; Institute of Experimental Physics, 040 01 Kosice, Slovakia} 
\affiliation{Joint Institute for Nuclear Research, RU-141980 Dubna, Russia} 
\affiliation{Duke University, Durham, North Carolina  27708} 
\affiliation{Fermi National Accelerator Laboratory, Batavia, Illinois 60510} 
\affiliation{University of Florida, Gainesville, Florida  32611} 
\affiliation{Laboratori Nazionali di Frascati, Istituto Nazionale di Fisica Nucleare, I-00044 Frascati, Italy} 
\affiliation{University of Geneva, CH-1211 Geneva 4, Switzerland} 
\affiliation{Glasgow University, Glasgow G12 8QQ, United Kingdom} 
\affiliation{Harvard University, Cambridge, Massachusetts 02138} 
\affiliation{Division of High Energy Physics, Department of Physics, University of Helsinki and Helsinki Institute of Physics, FIN-00014, Helsinki, Finland} 
\affiliation{University of Illinois, Urbana, Illinois 61801} 
\affiliation{The Johns Hopkins University, Baltimore, Maryland 21218} 
\affiliation{Institut f\"{u}r Experimentelle Kernphysik, Universit\"{a}t Karlsruhe, 76128 Karlsruhe, Germany} 
\affiliation{Center for High Energy Physics: Kyungpook National University, Daegu 702-701, Korea; Seoul National University, Seoul 151-742, Korea; Sungkyunkwan University, Suwon 440-746, Korea; Korea Institute of Science and Technology Information, Daejeon, 305-806, Korea; Chonnam National University, Gwangju, 500-757, Korea} 
\affiliation{Ernest Orlando Lawrence Berkeley National Laboratory, Berkeley, California 94720} 
\affiliation{University of Liverpool, Liverpool L69 7ZE, United Kingdom} 
\affiliation{University College London, London WC1E 6BT, United Kingdom} 
\affiliation{Centro de Investigaciones Energeticas Medioambientales y Tecnologicas, E-28040 Madrid, Spain} 
\affiliation{Massachusetts Institute of Technology, Cambridge, Massachusetts  02139} 
\affiliation{Institute of Particle Physics: McGill University, Montr\'{e}al, Canada H3A~2T8; and University of Toronto, Toronto, Canada M5S~1A7} 
\affiliation{University of Michigan, Ann Arbor, Michigan 48109} 
\affiliation{Michigan State University, East Lansing, Michigan  48824}
\affiliation{Institution for Theoretical and Experimental Physics, ITEP, Moscow 117259, Russia} 
\affiliation{University of New Mexico, Albuquerque, New Mexico 87131} 
\affiliation{Northwestern University, Evanston, Illinois  60208} 
\affiliation{The Ohio State University, Columbus, Ohio  43210} 
\affiliation{Okayama University, Okayama 700-8530, Japan} 
\affiliation{Osaka City University, Osaka 588, Japan} 
\affiliation{University of Oxford, Oxford OX1 3RH, United Kingdom} 
\affiliation{Istituto Nazionale di Fisica Nucleare, Sezione di Padova-Trento, $^x$University of Padova, I-35131 Padova, Italy} 
\affiliation{LPNHE, Universite Pierre et Marie Curie/IN2P3-CNRS, UMR7585, Paris, F-75252 France} 
\affiliation{University of Pennsylvania, Philadelphia, Pennsylvania 19104}
\affiliation{Istituto Nazionale di Fisica Nucleare Pisa, $^y$University of Pisa, $^z$University of Siena and $^{aa}$Scuola Normale Superiore, I-56127 Pisa, Italy} 
\affiliation{University of Pittsburgh, Pittsburgh, Pennsylvania 15260} 
\affiliation{Purdue University, West Lafayette, Indiana 47907} 
\affiliation{University of Rochester, Rochester, New York 14627} 
\affiliation{The Rockefeller University, New York, New York 10021} 
\affiliation{Istituto Nazionale di Fisica Nucleare, Sezione di Roma 1, $^{bb}$Sapienza Universit\`{a} di Roma, I-00185 Roma, Italy} 

\affiliation{Rutgers University, Piscataway, New Jersey 08855} 
\affiliation{Texas A\&M University, College Station, Texas 77843} 
\affiliation{Istituto Nazionale di Fisica Nucleare Trieste/Udine, $^{cc}$University of Trieste/Udine, Italy} 
\affiliation{University of Tsukuba, Tsukuba, Ibaraki 305, Japan} 
\affiliation{Tufts University, Medford, Massachusetts 02155} 
\affiliation{Waseda University, Tokyo 169, Japan} 
\affiliation{Wayne State University, Detroit, Michigan  48201} 
\affiliation{University of Wisconsin, Madison, Wisconsin 53706} 
\affiliation{Yale University, New Haven, Connecticut 06520} 
\author{T.~Aaltonen}
\affiliation{Division of High Energy Physics, Department of Physics, University of Helsinki and Helsinki Institute of Physics, FIN-00014, Helsinki, Finland}
\author{J.~Adelman}
\affiliation{Enrico Fermi Institute, University of Chicago, Chicago, Illinois 60637}
\author{T.~Akimoto}
\affiliation{University of Tsukuba, Tsukuba, Ibaraki 305, Japan}
\author{M.G.~Albrow}
\affiliation{Fermi National Accelerator Laboratory, Batavia, Illinois 60510}
\author{B.~\'{A}lvarez~Gonz\'{a}lez}
\affiliation{Instituto de Fisica de Cantabria, CSIC-University of Cantabria, 39005 Santander, Spain}
\author{S.~Amerio$^x$}
\affiliation{Istituto Nazionale di Fisica Nucleare, Sezione di Padova-Trento, $^x$University of Padova, I-35131 Padova, Italy} 

\author{D.~Amidei}
\affiliation{University of Michigan, Ann Arbor, Michigan 48109}
\author{A.~Anastassov}
\affiliation{Northwestern University, Evanston, Illinois  60208}
\author{A.~Annovi}
\affiliation{Laboratori Nazionali di Frascati, Istituto Nazionale di Fisica Nucleare, I-00044 Frascati, Italy}
\author{J.~Antos}
\affiliation{Comenius University, 842 48 Bratislava, Slovakia; Institute of Experimental Physics, 040 01 Kosice, Slovakia}
% ...

\collaboration{CDF Collaboration\footnote{With visitors from $^a$University of Massachusetts Amherst, Amherst, Massachusetts 01003,
$^b$Universiteit Antwerpen, B-2610 Antwerp, Belgium, 
$^c$University of Bristol, Bristol BS8 1TL, United Kingdom,
$^d$Chinese Academy of Sciences, Beijing 100864, China, 
$^e$Istituto Nazionale di Fisica Nucleare, Sezione di Cagliari, 09042 Monserrato (Cagliari), Italy,
$^f$University of California Irvine, Irvine, CA  92697, 
$^g$University of California Santa Cruz, Santa Cruz, CA  95064, 
$^h$Cornell University, Ithaca, NY  14853, 
$^i$University of Cyprus, Nicosia CY-1678, Cyprus, 
$^j$University College Dublin, Dublin 4, Ireland,
$^k$Royal Society of Edinburgh/Scottish Executive Support Research Fellow,
$^l$University of Edinburgh, Edinburgh EH9 3JZ, United Kingdom, 
$^m$Universidad Iberoamericana, Mexico D.F., Mexico,
$^n$Queen Mary, University of London, London, E1 4NS, England,
$^o$University of Manchester, Manchester M13 9PL, England, 
$^p$Nagasaki Institute of Applied Science, Nagasaki, Japan, 
$^q$University of Notre Dame, Notre Dame, IN 46556,
$^r$University de Oviedo, E-33007 Oviedo, Spain, 
$^s$Simon Fraser University, Vancouver, British Columbia, Canada V6B 5K3,
$^t$Texas Tech University, Lubbock, TX  79409, 
$^u$IFIC(CSIC-Universitat de Valencia), 46071 Valencia, Spain,
$^v$University of Virginia, Charlottesville, VA  22904,
$^{dd}$On leave from J.~Stefan Institute, Ljubljana, Slovenia, 
$^{ee}$Now at Freiburg University, Germany, 
}}
\noaffiliation

% ========================================================
% --------- End of author list and affiliations ----------
% ========================================================

\date{\today}

%
% ==============> Text of the abstract goes here <=====================
% ==============> Text of the abstract goes here <=====================
% 
\begin{abstract}
We report on a search for inclusive production of squarks and gluinos in $p \bar{p}$ collisions at 
$\sqrt{s}=1.96$~TeV, in events with large missing transverse energy and multiple jets of hadrons in
the final state. The study uses a CDF Run~II data sample corresponding to 2~fb${}^{-1}$ of integrated luminosity.
The data are in good agreement with the 
standard model predictions, giving no evidence for any squark or gluino component. 
In an R-parity conserving minimal supergravity scenario with $A_0 = 0$, $\mu < 0$ and tan$\beta = 5$,  95$\%$ C.L. upper limits on 
the production cross sections in the range between 0.1~pb  and 1~pb are obtained, depending on the squark and gluino masses considered. 
For gluino masses below 280 GeV/$c^2$, arbitrarily large squark masses are excluded at the 95$\%$ C.L., while 
for mass degenerate gluinos and squarks, masses below 392 GeV/$c^2$ are excluded at the 95$\%$~C.L.     
\end{abstract}

\maketitle

%\vfill\eject
%\twocolumn

% ====================
% ---- Introduction
% ====================

Supersymmetry (SUSY)~\cite{susy} is regarded as a possible extension of the standard model (SM) that 
naturally solves the hierarchy problem and provides a possible candidate for dark matter in the Universe.  SUSY introduces
a new symmetry that relates fermionic and bosonic degrees of freedom, and doubles the SM spectrum of particles by introducing a 
new supersymmetric partner (sparticle) for each particle in the SM.  
% The most general form of its minimal construction (MSSM)
% would allow a rapid proton decay. This is avoided by enforcing the conservation of an additional quantity, R-parity, 
% defined as $R = (-1)^{3(B-L)+2s}$, where $s$ denotes the spin and $L$ and $B$ are the lepton and baryon numbers, respectively. 
% As a consequence, in $R$-parity conserved scenarios, sparticles are produced in pairs and the lightest supersymmetric particle (LSP) 
% is stable. 
Results on similar inclusive searches for SUSY using Tevatron data have been previously reported by both the CDF and $\Dzero$ experiments 
in Run I~\cite{runI} and by the $\Dzero$ experiment in Run II~\cite{runII}.
This Letter presents new results on an inclusive search for squarks and gluinos, supersymmetric partners of quarks and gluons, based
on data collected by the CDF experiment in Run~II and corresponding to    
2.0 $\rm fb^{-1}$ of integrated luminosity.   
The analysis is performed within the framework of  minimal supergravity (mSUGRA)~\cite{msugra}  
and assumes R-parity conservation where sparticles are produced in pairs and the lightest supersymmetric 
particle (LSP) is stable, neutral, and weakly interacting. 
The expected signal is characterized by the production 
of multiple jets of hadrons from the cascade decays of squarks and gluinos and 
large missing transverse energy $\met$~\cite{coord} from the presence of two LSPs  in the final state. 
In a scenario with squark masses $\msq$ significantly larger than the gluino mass $\mgl$  at least four jets in the final state
are expected, while for  $\mgl > \msq$  dijet configurations dominate. 
Separate analyses are carried out for events with at least two, three, and four jets in the final state and with different requirements on the 
minimum $\met$.
The results are compared to SM background predictions from quantum chromodynamics (QCD) jets,  $W$ and $Z/\gamma^*$ bosons with 
accompanying jets ($W$+jets and $Z/\gamma^*$+jets), 
top quark, and diboson ($WW$, $ZW$ and $ZZ$) processes.

% ================
% ------ Detector
% ================

The CDF II detector is described in detail elsewhere~\cite{cdfii}.
The detector has a charged particle tracking system that is immersed in a 1.4~T solenoidal magnetic field 
coaxial with the beam line, and provides coverage in the 
pseudorapidity~\cite{coord} range $|\eta| \leq 2$. 
Segmented sampling calorimeters, arranged in a projective tower
geometry, surround the tracking system and measure the energy of 
interacting particles for $|\eta| < 3.6$. The central electromagnetic and hadronic calorimeters
cover the region $|\eta| < 1$,  
while the endwall hadronic calorimeter 
provides coverage out to 
$|\eta| < 1.3$. Forward  electromagnetic and hadronic 
calorimeters cover the regions $1.1 < |\eta| < 3.6$ and $1.3 < |\eta| < 3.6$, respectively.
Cherenkov counters in the region $3.7 < |\eta| < 4.7$ measure the number of 
inelastic $p \overline{p}$ collisions to determine the luminosity~\cite{clc}.

% ==============
% ---------- MC 
% ==============

Simulated event samples are used to determine detector acceptance and 
reconstruction efficiency, estimate SM background contributions, and compute the 
number of expected SUSY signal events.  Samples of simulated QCD-jets, $\tt$ production, and 
diboson processes are generated using the {\sc pythia} 6.216~\cite{pythia} 
Monte Carlo generator with Tune~A~\cite{underlying}. The normalization of the QCD-jets sample is extracted from data 
in a low $\met$ region, while $\tt$ and diboson samples are normalized to next-to-leading order (NLO) predictions~\cite{topnlo,dibosonnlo}.
Samples of simulated $Z/\gamma^*$+jets and $W$+jets events are
generated using the {\sc alpgen} 2.1 program~\cite{alpgen} where exclusive 
subsamples with different jet multiplicities are combined, and the resulting samples are  
normalized to the measured $Z$ and $W$ inclusive cross sections~\cite{dy}.
Finally, samples of single top  events are produced using the {\sc madevent} program~\cite{madevent} 
and normalized using NLO predictions~\cite{stnlo}. 
% CTEQ5L~\cite{cteq} parton distribution functions (PDFs) 
% are used for the proton and antiproton and a value of 172~GeV/$c^2$ is used.
%
In mSUGRA, the mass spectrum of sparticles is determined by five parameters: 
the common scalar and gaugino masses at the GUT scale, $M_0$ and $M_{1/2}$, respectively; 
the common trilinear coupling at the GUT scale, $A_0$; the sign of the Higgsino mixing parameter, $\mu$; 
and the ratio of the Higgs vacuum expectation values, tan$\beta$.    
The mSUGRA samples are generated using the {\sc isasugra} implementation in  
{\sc pythia} with $A_0 = 0$, $ \mu < 0$, and tan$\beta = 5$, as inspired by previous studies~\cite{snow}. A total of 132 different
squark and gluino masses are generated via variations of $M_0$ and $M_{1/2}$ in the range 
$M_0 < 600$~GeV/$c^2$ and $50 < M_{1/2} < 220$~GeV/$c^2$.  At low  tan$\beta$, the squarks 
from the first two generations are nearly degenerate, whereas the mixing of the third generation 
leads to   slightly lighter sbottom masses 
and  much lighter stop masses. In this analysis, stop pair production processes are not considered.   
The contribution from hard processes involving sbottom production is almost negligible, 
and is not included in the calculation of the signal efficiencies 
 to avoid a dependency on the details of the model for squark mixing. 
The mSUGRA samples are 
normalized using NLO cross sections as determined by {\sc prospino} 2.0~\cite{prospino}, with input parameters provided
by {\sc isajet} 7.74~\cite{isajet}. CTEQ61M  parton distribution functions (PDFs)~\cite{cteq} are used, and renormalization 
and factorization scales are set to the average  mass~\cite{mass} of the sparticles produced in the hard interaction.   
The Monte Carlo events are passed through a full
CDF II detector simulation (based on  {\sc geant3}~\cite{geant} and {\sc gflash}~\cite{gflash})
% package is used to simulate the energy deposition in the calorimeters) 
and reconstructed 
and analyzed with the same analysis chain as for the data.

%========================
% ---- Event Selection
% =======================

Data are collected using a three-level trigger system that selects events with $\met > 35$~GeV and 
at least two 
calorimeter clusters with  $E_T$ above 10~GeV.
The events are then required to have a primary vertex
with a $z$ position within 60~cm of the nominal interaction. 
Jets are reconstructed from the energy deposits in the calorimeter towers using a cone-based jet
algorithm~\cite{jetclu} with cone radius $R=\sqrt{\Delta\eta^2 + \Delta\phi^2 } = 0.7$,
and the measured $\etjet$ is corrected for detector effects and contributions from 
multiple $p\bar{p}$ interactions per crossing at high instantaneous luminosity, as
discussed in Ref.~\cite{jetcornim}. The events are required to have 
at least two, three, or four jets (depending on the
final state considered), each jet with corrected transverse energy 
$\etjet > 25$~GeV and  pseudorapidity  in the range $|\etajet|<2.0$, and 
at least one of the jets is required to have  $|\etajet|<1.1$.
Finally, the events are  
required to have $\met > 70$~ GeV.
For the kinematic range in $\met$  and the $\etjet$ of the jets considered in this analysis,  
the trigger selection is 100$\%$ efficient. 
Beam-related backgrounds and cosmic rays are removed by  
requiring an average jet electromagnetic fraction 
$ f_{\rm em} = \sum_{\rm jets} \etjetem / \sum_{\rm jets} \etjet > 0.15$,
where $\etjetem$ denotes the electromagnetic component of the jet transverse energy,  
and the sums run over all the selected jets in the event. 
In addition, the events are required to have an average charged particle
fraction $ f_{\rm ch} = \sum_{\rm jets} \ptch / \sum_{\rm jets} \etjet > 0.15$, where, for each selected 
jet with $|\etajet| < 1.1$, 
$\ptch$ is computed as the scalar sum of the transverse momenta  $p_T^{\rm track}$  of tracks with 
$p_T^{\rm track} > 0.3$~GeV/$c$ 
and within a cone of radius $R=0.4$ around the jet axis.
The requirements on $f_{\rm em}$ and $ f_{\rm ch}$ reject events with anomalous energy deposition in the hadronic 
section of the calorimeter or energy deposits in the calorimeter inconsistent with the observed activity in the tracking system, 
 and have no significant effect on the mSUGRA and SM Monte Carlo samples.

The dominant QCD-jets background with large $\met$  originates from the misreconstruction of the jet 
energies in the calorimeters. In such events the $\met$ direction tends to be  aligned, in the transverse plane, with one of the leading 
jets in the event. This background contribution is suppressed by requiring an azimuthal separation  
$\Delta \phi (\met - \rm jet) > 0.7$ for each of the selected jets in the event. In the case of the four-jets 
analysis, the requirement for the least energetic jet is limited to $\Delta \phi (\met - \rm jet) > 0.3$.
Finally, in the two-jets analysis case the events are rejected if they contain 
a third jet 
with $\etjet > 25$~GeV, $|\etajet|<2.0$, and $\Delta \phi (\met - \rm jet) < 0.2$. 
%
% These selection criteria remove most of the QCD-jets 
% background, for which large $\met$  originates from the misreconstruction of the jet energies in the calorimeter towers, and the $\met$
% direction tends to be  aligned, in the transverse plane, to  one of the leading jets in the event.
%
%
The SM background contributions with energetic 
electrons~\cite{electron} in the final state from $Z$ and $W$ 
decays are suppressed by 
requiring $ \etjetem / \etjet < 0.9$ for each selected jet in the event.  In addition, events that have one isolated track with 
$p_T^{\rm track} > 10$~GeV/$c$ and $\Delta \phi (\met - \rm track) < 0.7$, or two isolated tracks 
with an invariant mass  $76 < M_{\rm trks} < 106$~GeV/$c^2$,  are vetoed to reject backgrounds with 
$W$ or $Z$ bosons decaying into muon or tau leptons. 

% =======================
% ------ Optimization
% =======================

% Different final states with at least two, three or four jets are considered  
% with the aim to increase the search sensitivity in a wide range of squark and gluino masses.
An optimization is carried out to determine, for each final state, 
the lower thresholds on $\met$, the $\etjet$ of the individual jets, 
and $H_T$, defined as $ H_T = \sum_{\rm jets} \etjet$~\cite{ht}. 
For each mSUGRA sample,  
the procedure  
maximizes $S/\sqrt{B}$, where $S$ denotes the number of SUSY
events and  $B$ is the total SM background. 
The results from the different mSUGRA samples are then combined to define, for each final state, a single set 
of lower thresholds 
that maximizes the 
search sensitivity in the widest range of squark and gluino masses (see Table~I). 
As an example, for $\msq = \mgl$ and masses between  300~GeV/$c^2$ and 400~GeV/$c^2$, values for $S/\sqrt{B}$
in the range between 20 and 6 are obtained, corresponding to SUSY selection efficiencies of 4$\%$ to 12$\%$, respectively.

\begin{table}[htbp]
\begin{footnotesize}
\begin{center}
\renewcommand{\baselinestretch}{1.2}
\begin{tabular}{|c c c c c c c|} \hline
\multicolumn{7}{|c|}{\small{Lower thresholds (GeV)}} \\ \hline\hline
  {\rm Final state}              & $\met$ & $H_T$ & $\etjeta$ & $\etjetb$ & $\etjetc$ & $\etjetd$ \\ 
$\met + \geq 2$ jets   & 180     & 330  & 165       &   100     &   -       &    -      \\ 
$\met + \geq 3$ jets   & 120     & 330  & 140       &   100     &  25       &    -      \\ 
$\met + \geq 4$ jets   &  90     & 280  &  95       &    55     &  55       &    25     \\ \hline
 \end{tabular}
\end{center}
\renewcommand{\baselinestretch}{1}
\end{footnotesize}
\label{tab:table1}
\caption{
Optimized lower thresholds on $\met$, $H_T$, and  $\etjeti \ (i=1-4)$ for each analysis.
}
\end{table}

% ===========================
% -------- control samples
% ============================

A number of control samples in data are considered to test the validity of the SM background predictions,
as extracted from simulated events.
The samples are defined by reversing the logic of some of the selection criteria described above. A sample dominated by QCD jets 
is obtained by requiring that at least one of the selected jets is aligned with the $\met$ direction.
A control sample dominated by $Z/\gamma^*$+jets, $W$+jets, and top-quark processes with highly energetic electrons in the final state is
obtained after requiring 
$ \etjetem / \etjet > 0.9$ 
for at least one of the jets. Similarly, a control sample with highly energetic
muons in the final state  is created
by requiring the presence of an isolated track with  $p_T^{\rm track} > 10$~GeV/$c$ and $\Delta \phi (\met - \rm track) < 0.7$, or two isolated
tracks with $76 < M_{\rm trks} < 106$~GeV/$c^2$. Good agreement is observed between the data and the SM 
predictions in each of the control regions for all the final states considered.

% =================================
% ---- systematic uncertainties
% =================================

A detailed study of the systematic uncertainties 
% on the estimation of SM background contributions 
% and mSUGRA signal efficiencies 
is  carried out for each final state~\cite{xavier,gdl}.  
A $3 \%$ uncertainty on the absolute jet energy scale~\cite{jetcornim} in the calorimeter introduces
an uncertainty in the background prediction that varies between 
$24 \%$ and $34 \%$, and an uncertainty on the mSUGRA
signal efficiencies between $15 \%$ and $17\%$. Uncertainties related to the modeling of
the initial- and final-state soft gluon radiation in the simulated samples translate into a $3\%$ to
$6\%$ uncertainty on the mSUGRA signal efficiency, 
and uncertainties on the background predictions that vary between $8\%$ and $10\%$.  
An additional $10 \%$ uncertainty on the diboson and top quark contributions accounts for the uncertainty on 
the predicted cross sections at NLO. A $2\%$ uncertainty on the measured Drell-Yan cross sections, 
relevant for $Z/\gamma^*$+jets and $W$+jets processes, is also included.
% whereas the normalization 
% of the QCD-jets Monte Carlo samples is determined from data.  
The total systematic uncertainty on the SM predictions varies between
$31 \%$ and $35 \%$ as the jet multiplicity increases.      
%% Finally, a $5.8 \%$ uncertainty on the total 
%% luminosity is included in the measured cross sections.
Various sources of uncertainty in the mSUGRA cross sections at NLO, as determined using {\sc prospino}, are considered. 
The uncertainty on the PDFs 
%is computed using the Hessian method~\cite{hessian,90cl}. It 
varies
between $10 \%$ and $20 \%$, depending on the mSUGRA point considered. Variations of the renormalization and factorization 
scales by a factor of two change the theoretical cross sections by $20 \%$ to $25 \%$.      

% ===============
% ---- results
% ===============

\begin{figure}
\includegraphics[width=\figsize]{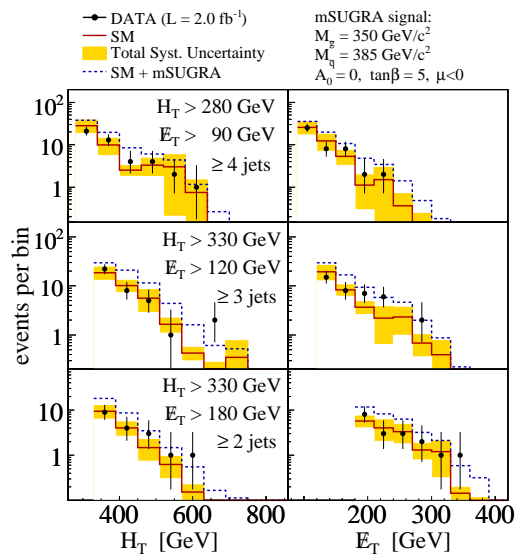}
\caption{ Measured $H_T$ and $\met$ distributions (black dots) in events with at least two (bottom), three (middle), 
and four (top) jets in the final state compared to 
the SM predictions (solid lines) and the SM+mSUGRA predictions (dashed lines).   
The shaded bands show the total systematic uncertainty on the SM predictions. 
}
\label{fig:metht}
\end{figure}

Figure~\ref{fig:metht} shows the measured $H_T$ and $\met$ distributions 
compared to the SM predictions after all final selection criteria are applied.  
For illustrative purposes, 
the Figure indicates the impact of a given mSUGRA 
scenario. The measured distributions are in good agreement with the SM predictions
in each of the three final states considered.
In Table~II, the observed number of events and the SM predictions are presented for each final state. 
A global $\chi^2$ test, including correlations between systematic uncertainties, gives 
a $94 \%$ probability.

\begin{table}[htbp]
\begin{footnotesize}
\begin{center}
\renewcommand{\baselinestretch}{1.2}
\begin{tabular}{|c c c c|} \hline
\multicolumn{4}{|c|}{\small{Events in data (2 fb${}^{-1}$)}} \\ \hline\hline
                                       & $\geq 2$ jets &  $\geq 3$ jets & $\geq 4$ jets \\
                                       &        18           &       38             &      45             \\ \hline\hline
\multicolumn{4}{|c|}{\small{SM  predictions}} \\ \hline\hline
% SM process                            & $\geq 2$ jets   & $ \geq 3$ jets  & $ \geq 4$ jets  \\
QCD jets                               &$      4.4\pm{}2.0    $&$     13.3\pm{}4.6     $&$    15.3\pm{}7.1    $\\
top                                    &$      1.3\pm{}1.2    $&$      7.6\pm{}4.1     $&$    22.1\pm{}7.0    $\\
$Z \rightarrow \nu \bar{\nu}$+jets     &$      3.9\pm{}0.9    $&$      5.4\pm{}1.4     $&$     2.7\pm{}0.7    $\\
$Z/\gamma^* \rightarrow l^+l^-$+jets   &$      0.1\pm{}0.1    $&$      0.2\pm{}0.1     $&$     0.1\pm{}0.1    $\\
$W \rightarrow l \nu$+jets             &$      6.1\pm{}2.2    $&$     10.7\pm{}3.1     $&$     7.7\pm{}2.2    $\\
$WW,ZW,ZZ$                             &$      0.2\pm{}0.2    $&$      0.3\pm{}0.2     $&$     0.5\pm{}0.2    $\\ \hline
total SM                               &$       16\pm{}5      $&$       37\pm{}12      $&$      48\pm{}17     $\\ \hline\hline
 \end{tabular}
\end{center}
\renewcommand{\baselinestretch}{1}
\end{footnotesize}
\label{tab:table2}
\caption{
Number of events in data for each final state compared to SM  predictions, including statistical and
systematic uncertainties summed in quadrature.
}
\end{table}

% =================
% ----- limits
% =================

\begin{figure}[h]
\includegraphics[width=\figsize]{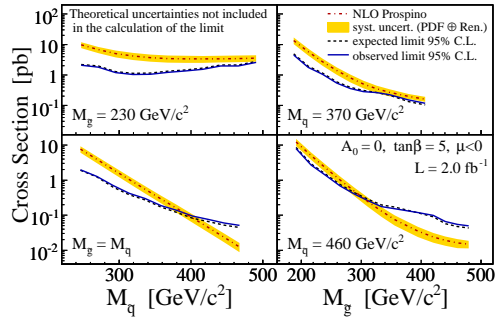}
\caption{Observed (solid lines) and expected (dashed lines) 95$\%$ C.L. upper limits on the inclusive squark and gluino production 
cross sections as a function of $\msq$(left) and $\mgl$(right) in different regions of the 
squark-gluino mass plane, compared to NLO mSUGRA predictions (dashed-dotted lines). The shaded bands denote 
the total uncertainty on the theory.
}
\label{fig:glu2}
\end{figure}

\begin{figure}[h]
\includegraphics[width=\figsizes]{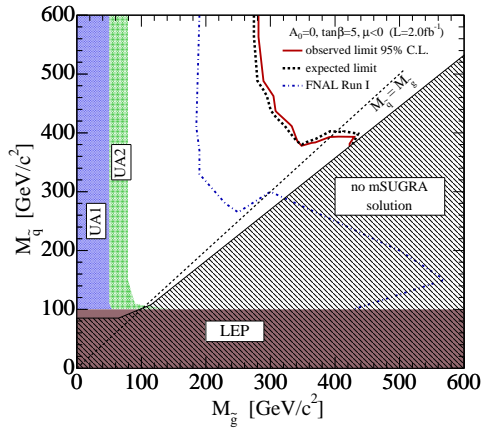}
\caption{Exclusion plane at 95 $\%$ C.L. as a function of squark and gluino masses in an mSUGRA scenario with 
$A_0 = 0$, $\mu < 0$ and tan$\beta = 5$. The observed (solid line) and expected (dashed line) upper limits are compared to 
previous results from SPS~\cite{ua12} and  LEP~\cite{lep} experiments at CERN (shaded bands), and from the Run I at the Tevatron~\cite{runI} 
(dashed-dotted line). The hatched area indicates the region in the plane with no mSUGRA solution. 
}
\label{fig:glu3}
\end{figure}

The results are translated into 95$\%$ C.L. upper limits on the cross section for squark and gluino production 
in different regions of the squark-gluino mass plane, using a Bayesian approach~\cite{bayes} and including
statistical and systematic uncertainties. For the latter, correlations between systematic uncertainties on signal efficiencies 
and background predictions are taken into account, and an additional $6 \%$ uncertainty on the total luminosity 
is included.  For each mSUGRA point considered, observed and expected limits are computed separately for each of the three 
analyses, and the one with the 
best expected limit is adopted as the nominal result. Figure~\ref{fig:glu2} shows the observed and expected  95$\%$ C.L. upper 
limits as a function of squark and gluino masses, compared to mSUGRA  predictions, in four  regions 
within the squark-gluino mass plane. 
Cross sections in the range between 0.1~pb and  1~pb are excluded by this analysis, depending on the 
masses considered. The observed numbers of events in data are also  translated into 95$\%$ C.L. upper limits for squark and 
gluino masses, for which the uncertainties on the theoretical cross sections are included in 
the limit calculation, and where the three analyses are combined in a similar way as for the cross section limits. 
Figure~\ref{fig:glu3} shows the excluded region in the squark-gluino mass plane. For the 
mSUGRA scenario considered, all squark masses 
are excluded for $\mgl < 280$~GeV/$c^2$, while for  $\msq =  \mgl$ masses up to 392~GeV/$c^2$ are excluded.
Finally, for  $\msq < 400$~GeV/$c^2$ gluinos with $\mgl < 340$~GeV/$c^2$ are excluded. This analysis 
extends the previous Run~I limits from the Tevatron  
by 80~GeV/$c^2$ to 140~GeV/$c^2$.

% ==================
% ---- conclusions
% ==================

In summary, we report results on an inclusive search for squarks and gluinos 
in  $p \overline{p}$ collisions at $\sqrt{s} = 1.96 \ {\rm TeV}$ in events with 
large $\met$ and multiple jets in the final states,  
based on 2~fb${}^{-1}$ of CDF Run II data. The measurements are in good agreement with SM predictions for backgrounds. 
The results are translated into 95$\%$ C.L. upper limits on production 
cross sections and squark and gluino masses in a given mSUGRA scenario, which significantly extend 
Run I results.  

%\begin{center}
%\textbf{Acknowledgments}
%\end{center}

We thank the Fermilab staff and the technical staffs of the participating institutions for their vital 
contributions. This work was supported by the U.S. Department of Energy and National Science Foundation; 
the Italian Istituto Nazionale di Fisica Nucleare; the Ministry of Education, Culture, Sports, Science and 
Technology of Japan; the Natural Sciences and Engineering Research Council of Canada; the National Science Council 
of the Republic of China; the Swiss National Science Foundation; the A.P. Sloan Foundation; the 
Bundesministerium f\"ur Bildung und Forschung, Germany; the Korean Science and Engineering Foundation and the 
Korean Research Foundation; the Science and Technology Facilities Council and the Royal Society, UK; the Institut 
National de Physique Nucleaire et Physique des Particules/CNRS; the Russian Foundation for Basic Research; the 
Ministerio de Ciencia e Innovaci\'{o}n, and Programa Consolider-Ingenio 2010, Spain; the Slovak R\&D Agency; and the 
Academy of Finland.


\begin{thebibliography}{99}
\bibitem{coord} CDF uses a cylindrical coordinate system about the beam axis 
with polar angle $\theta$ and azimuthal angle $\phi$.
We define  transverse energy $E_T = E \ {\rm sin} \theta$, transverse momentum  
$p_T = p \ {\rm sin}  \theta$, pseudorapidity $\eta = - {\rm ln(tan(\frac{\theta}{2})) }$, and rapidity  
$y = \rm \frac{1}{2} ln(\frac{E+p_z}{E-p_z}) $. The missing transverse energy $\met$ is defined as the norm of $-\sum_i E^i \cdot \vec{n}_i$, 
where $\vec{n}_i$ is the component in the azimuthal plane of the unit vector pointing
from the interaction point to the i-th calorimeter tower.
\bibitem{mass} Pole masses are considered. The squark mass is averaged over the first two squark generations.
\bibitem{electron} Charge conjugation is implied throughout the paper.
\bibitem{ht} The sum runs over the selected jets. In the four-jets case, the first three leading jets are considered.
\end{thebibliography}
\end{document}